# The self-compression of injected electron-hole plasma in silicon

P. D. Altukhov* and E. G. Kuzminov

A. F. Ioffe Physical-Technical Institute, Politekhnicheskaya street 26,

St. Petersburg 194021, Russia

*Corresponding author: P. D. Altukhov; Tel: +7-812- 292-7344;
Fax: +7-812-297-1017;
E-mail address: pavel.altukhov@gmail.com

**Abstract.** A recombination radiation line of electron-hole plasma, observed in electroluminescence spectra of tunneling silicon MOS diodes, has been investigated at the temperature $T \geq 300\,K$. The internal quantum efficiency of the luminescence, equal to $(1 \div 3) \times 10^{-3}$, appears to be unexpectedly high. The spectral position of the luminescence line indicates, that a weak overheating of the diode by the diode current results in an anomalously strong reduction of the semiconductor energy gap inside the electron-hole plasma. A unique threshold optical hysteresis is observed in the luminescence intensity with changing diode current. These results are explained by condensation of injected electron-hole plasma into a dense state. A reduction of the semiconductor energy gap due to generation of phonons by the plasma is discussed as a reason of the plasma condensation. The plasma condensation is identified as the plasma self-compression.

**Introduction**

The problem of silicon optoelectronics, important for semiconductor physics and engineering, was unresolved in the last thirty years of the twentieth century. Weak radiative recombination and strong nonradiative Auger recombination restrict the quantum efficiency of silicon luminescence at room temperature. At low temperatures recombination radiation lines of free excitons [1] and the electron-hole liquid [2] dominate in luminescence spectra of pure silicon. The quantum efficiency of luminescence of free excitons in pure silicon is $\eta \sim (10^{-3} \div 10^{-2})$ [2], where $\eta = \tau_R/\tau_O$, $\tau_O$ is the recombination radiation time, $\tau_R$ is the recombination time.

At $T \geq 300\,K$ luminescence in silicon is represented by recombination radiation of free electrons and holes [3,4]. In accordance with the theory of Roosbroek and Shockley [5] the recombination radiation time of free electrons and holes in silicon, corresponding to a measured absorption coefficient for indirect optical transitions [6], can be represented for the most intensive TO-line as $\tau_o \approx [A_o n(1+N_q)]^{-1}$ [7]. Here $n$ is the density of electron-hole pairs, $A_o \approx 3 \times 10^{-15}\,cm^3 s^{-1}$, $N_q$ is the filling number of the TO-phonon. $N_q$ is much lower than unit at $T < 500\,K$. The quantum efficiency of room-temperature luminescence of electron-hole plasma in silicon for the most intensive TO-line can be expressed as

$$\eta \approx A_o n(1+N_q)(\tau_{R0}^{-1} + G_A n^2)^{-1}. \tag{1}$$

This gives the maximum value of the quantum efficiency $\eta \approx 5 \times 10^{-3}$ at the pair density $n \approx 1.3 \times 10^{18}\,cm^{-3}$. Here $\tau_{R0} \approx 2.5 \times 10^{-6}\,s$ is the recombination time at low densities of electron-hole pairs and $G_A \approx 2.3 \times 10^{-31}\,cm^6 s^{-1}$ is the coefficient of Auger recombination in silicon [8]. These estimates show, that a high quantum efficiency of silicon luminescence, available for applications in optoelectronics, can be achieved at room temperature. Nevertheless, during a long time studied silicon room-temperature luminescence was weak ($\eta < 10^{-5}$) [4,9]. An erbium-doped silicon light emitting diode gives room-temperature luminescence with $\eta \approx 10^{-4}$ [10]. Observation of photoluminescence of large silicon clusters in a silicon oxide matrix with the external quantum efficiency grater than $10^{-3}$ [11] represents an achievement for special silicon-based structures.

A new surprising phenomenon – condensation of injected electron-hole plasma into dense plasma flexes or into a dense surface state, observed in tunneling silicon MOS structures at $T > 300 K$ [7], results in the maximum quantum efficiency of electroluminescence in silicon $\eta \sim (10^{-3} \div 10^{-2})$. Intense room-temperature electroluminescence in similar tunneling silicon MOS structures was investigated by Liu *et al* [12]. An efficient room-temperature silicon light emitting diode with the quantum efficiency $\eta \geq 10^{-3}$ has been realized recently by use of a dislocation rich p-n junction [13].

Here we represent new experimental results supporting our conclusion [7] about high temperature condensation of injected electron-hole plasma in silicon. New investigations of the recombination radiation line of electron-hole plasma in electroluminescence spectra of tunneling silicon MOS diodes were performed at the lattice temperature a little bit higher than room temperature under an effective cooling of the diodes. A remarkable threshold optical hysteresis, observed in the luminescence intensity with changing diode current, represents a strong evidence of the plasma condensation. A simple theoretical model of a plasma-phonon condensation, based on a negative heat capacity of the electron-hole plasma at high temperatures, is introduced as an explanation of our results.

## Results and discussion

**Tunneling silicon MOS diodes**

Tunneling silicon MOS diodes (figure 1) were fabricated on n-type silicon substrates with the phosphorus concentration $n_D \approx 3 \times 10^{14} cm^{-3}$ and on p-type silicon substrates with the boron concentration $n_A \approx 8 \times 10^{14} cm^{-3}$. The highly doped diffusion contact was made on the back surface of the substrate. The substrate thickness is $d \approx 3 \times 10^{-2} cm$. The diameter of the tunneling area is $5 \times 10^{-2} cm$. The thickness of the SiO$_2$ oxide in the tunneling area is $d_o \approx 10 A°$. The SiO$_2$ oxide with the thickness $2000 A°$ was grown by the dry oxidization at $960°C$. The tunneling area was made by a selective etching of the oxide. The oxide is covered by the NiCr film with the thickness $300 A°$. The top Al gate contact with the thickness higher than $10^{-4} cm$ is used. The diode was mounted on a copper plate for a diode cooling. Sizes of the copper plate were changed for a variation of a cooling regime. An electrical and a thermal contact of the back surface of the silicon substrate and the copper plate were realized by use of a conductive silver paste. Recombination radiation from the diode was collected through an aperture in the copper plate and analyzed by a spectrometer. The light intensity was detected by a photomultiplier with the photocathode of the *S*-1 type, operating at $T = 80 K$. A spectral distribution of the spectrometer sensitivity was used in experiments for a correction of recombination radiation spectra.

Elecroluminescence spectra of two-dimensional carriers at a silicon surface in the diodes at $T = 1.5 K$ [14-16] show, that the surface quantum well in the diodes is deep and screening of an electrical field at a silicon surface should be almost quantum at room temperatures and densities of surface carriers $n_S \approx (1 \div 3) \times 10^{13} cm^{-2}$. However, the density of carriers at upper quantum levels in the surface potential well can be relatively high at room temperatures. In the case of the Si:P diode (figure 1) this can give a tail in the surface potential, which is repulsive for holes. The additional repulsion of holes from the surface could prevent the formation of the self-organized hole potential well, observed in experiments on optical spectroscopy of two-dimensional electrons [17-20]. The surface hole potential well is not essential for interpretation of our results, and we have not shown it in figure 1.

The tunneling silicon MOS diode is a convenient device for realization of a strong and effective injection of nonequilibrium carriers. In the case of the Si:P diode the tunneling current of holes $J_h$ represents at high temperatures the bulk injection current, equal to the bulk recombination current $J_R$ (figure 1). Hence, $J_h = J_R$, where the bulk recombination current is $J_R = e n_P \tau_R^{-1}$, $n_P = \int n_h dz$ is the number of injected holes per square centimeter, $n_h$ is the density of injected nonequilibrium holes in the silicon substrate. The efficiency of the injection is $\gamma = J_h / J$, where $J = J_h + J_e + J'$ is the diode current, $J_e$ is the tunneling current of surface electrons, and $J'$ is the tunneling current of surface states in the semiconductor energy gap [7]. The diode efficiency is $\gamma \eta$. A high tunneling current of holes, defined by the injection potential $U'$, and a high efficiency of injection can be achieved at relatively low values of $U'$ and low energies of injected holes, if the oxide thickness is low [16].

**The diffusion-drift model**

The problem of distribution of injected carriers in the silicon substrate is the basic problem of our work. It is obvious, that injected electron-hole plasma should be almost neutral and $n_e \approx n_h = n$, where $n_e$ is the density of injected nonequilibrium electrons, $n$ is the density of injected electron-hole plasma. A space charge should be concentrated at the silicon surface. At a low plasma density the mobility of electrons and holes is defined by electron-phonon scattering and the hole mobility is lower than the electron mobility [21-23]. Hence, in the Si:P diode at a low plasma density the drift velocity of injected holes $v_h = \mu_h \mathcal{E}$ is lower than the drift velocity of electrons $v_e = \mu_e \mathcal{E}$, and electrons give the main contribution to the diode current. Here $\mu_h, \mu_e$ are the hole and electron mobilities, $\mathcal{E}$ is the electrical field. The simplest model for distribution of injected electron-hole plasma in the Si:P diode is the diffusion-drift model. In accordance with this model the density of injected electron-hole plasma at the silicon surface decreases exponentially with increasing distance from the surface, the average distance of injected carriers from the surface is $L_D = (D\tau_R)^{1/2}$ and $n_P = nL_D$. Here $D$ is the diffusion coefficient, $L_D$ is the diffusion length. An electrical field in the plasma and the drift velocities of electrons and holes increase with increasing distance. At a large distance in a strong electrical field the drift currents of electrons and holes can be higher than the diffusion currents, and a drift regime of low-density plasma (the system of free electrons and holes) can be achieved. This solution can be obtained from the continuity equations for electrons and holes [23]. At room temperature and $\tau_R \approx 10^{-6} s$ the diffusion length is $L_D \approx 3 \times 10^{-3} cm$. The luminescence intensity $I$ of the diode is proportional to the radiative recombination current $J_R^r = \gamma \eta J$. At low diode currents and the injection efficiency close to unit the luminescence intensity in this model depends on the diode current as $I \sim J^2$. At high diode currents, when the Auger recombination becomes essential, this dependence transforms itself into $I \sim J^{1/2}$ and it does not include a long linear part. Experimental results, represented in this paper, and our previous results [7] do not agree to the diffusion-drift model.

Band diagram in the case of the Si:B diode [7], corresponding to the negative sign of the gate potential $V_g$ and the substrate potential $V_s$, includes the p$^+$-contact instead of the n$^+$-contact, shown in figure 1, and the hole surface potential well instead of the electron surface potential well. In the Si:B diode at a low plasma density the drift velocity of injected electrons is higher than the drift velocity of holes, but electrons can give the main contribution to the diode current only at $n > n_A$. At the same time the drift length of electrons $v_e \tau_R$ and the drift length of holes $v_h \tau_R$ in a sufficiently strong electrical field can be higher than the diffusion length and the substrate thickness. The simplest model for distribution of injected electron-hole plasma in the Si:B diode, operating in a strong electrical field [7], is the drift model. In accordance with this model in a drift regime of the plasma $v_e \tau_R$, $v_h \tau_R \gg d \gg L_D$, the diffusion currents of electrons and holes in the silicon substrate are weak and the recombination current is $J_R = end\tau_R^{-1} + en_{PS}\tau_{RS}^{-1}$. Here $n_{PS}$ is the number of injected carriers at the p$^+$-contact and $\tau_{RS}$ is the recombination current at the p$^+$-contact. Variation of the plasma density with increasing distance from the surface is weak. The drift model corresponds to a weak recombination current in the silicon substrate, a strong nonradiative recombination current at the p$^+$-contact and a low efficiency of the diode. It is evident, that a proper diffusion-drift regime is organized for electron-hole plasma at the p$^+$-contact. At room temperature conditions of a drift regime in our diodes can be realized in the electrical field $\mathcal{E} \geq 300\, V\, cm^{-1}$ corresponding to the gate potential $|V_g| \geq 10\, V$. Experimental results, represented in this paper, and our previous results [7] do not agree to the drift model.

At room temperature and the plasma density $n \geq 10^{18}\, cm^{-3}$ electron-hole scattering gives the main contribution to the mobility of electrons and holes, and the scattering length of electrons and holes is comparable with the average distance between carriers $n^{-1/3}$. The probability of electron-hole scattering can be approximately described by the formula of Brooks and Herring [21-23]. The probability of electron-electron scattering, the probability of hole-hole scattering and the probability of inelastic scattering of electrons and holes should be very close to the probability of electron-hole scattering. In a strong electrical field the probability of electron-hole scattering decreases due to an increase of the electronic temperature. At the electronic temperature $T_e \geq 600\, K$ emission of optical phonons by hot electron-hole plasma in a strong electrical field should give a decrease of the mobility of electrons and holes [22,23].

So, analysis of the continuity equations indicates, that injected plasma can be concentrated at the silicon surface or at the metal contact in a diffusion-drift regime. This gives a corresponding concentration of the recombination current. In both cases at a large distance from the surface or from the metal contact a drift regime of the injected plasma can be achieved in a strong electrical field. The electrical field in dense surface plasma should be weak due to screening. A decrease of the plasma density results in a corresponding increase of the electrical field. This keeps the diode current independent from the distance.

**Experiment**

In the case of the Si:P diodes, the strongly nonlinear voltage-current dependence of the diode (figure 2) shows, that tunneling injection produces a strong modulation of the substrate conductivity and the diode current $J = e[v_e(n_D + n) + v_h n]$, where $J = J°S^{-1}$, $J°$ is the total current of the diode, $S \approx 2\times10^{-3} \, cm^2$ is the area of the diode. At the maximum diode current $J = 100 \, A \, cm^{-2}$ the plasma density near the n$^+$-contact is estimated as $n \geq n_D$. This implies, that a drift regime of low-density plasma is achieved at high diode currents and a large distance from the surface. At $J° \geq 40 \, mA$ the tunneling resistance of the oxide is low at a low injection potential $U'$ (figure 1) and the resistance of the diode is close to the resistance of the silicon substrate. Here we use an effective cooling of the diodes, supplying weak overheating of the lattice and the intrinsic concentration of carriers in the silicon substrate lower than the donor or acceptor concentration. In previous experiments [7] the diode cooling was not effective. It has been found, that with increasing cooling and decreasing diode temperature the luminescence intensity decreases due to a shift of the luminescence threshold in the intensity-current dependence to the region of high diode currents and due to a decrease of the linear part of this dependence. At weak diode cooling the linear part of the intensity-current dependence is very long [7]. The voltage-current dependence of the Si:P diode is reversible (figure 2), but the intensity-current dependence reveals a well-defined hysteresis. The luminescence intensity for increasing diode current (figure 2, curve $I_+$) is lower than the luminescence intensity for decreasing diode current (figure 2, curve $I_-$). The diode current was changed with the rate $2.5\times10^{-5} \, A \, s^{-1}$, supplying a complete temperature relaxation of the diode. It is difficult to explain the intensity-current dependence in the diffusion-drift model.

The TO-line, corresponding to emission of the TO-phonon in the recombination process, gives the main contribution to luminescence spectra of free carriers. The intensity of the TA-line is by the order of magnitude lower than the intensity of the TO-line. The EH-line in the luminescence spectra (figure 3) is attributed to recombination radiation of free electrons from the indirect conduction band minima and free holes from the valence band $\Gamma_8$. The line shape of the EH-line in the case of the Si:P diode is described by the formula $I = I_o E^2 \exp(-E/kT')$, where $T' = \varsigma T_e$, $E = h\nu - h\nu_o$, $h\nu_o = E_g - \hbar\Omega^{TO}$, $E_g$ is the energy gap of silicon, $\hbar\Omega^{TO} = 58\ meV$ is the energy of the TO-phonon. A difference between $T'$ and $T_e$ is explained by absorption of the light in the silicon substrate [7]. Estimates by use of the absorption coefficient in silicon [6] give for our diodes at room temperature $\varsigma \approx 0.8$, if the electron-hole plasma is concentrated at the silicon surface, and $\varsigma \approx 0.9$, if the electron-hole plasma is distributed uniformly in the silicon substrate. Therefore, measurements of the spectral temperature $T'$ give a reasonable estimate of the electronic temperature $T_e$ of the electron-hole plasma. The theoretical line shape coincides with the experimental line shape except the low energy and the high energy tails of the EH-line (figure 3). A significant contribution to the low energy tail can be given presumably by a two-phonon emission line of the electron-hole plasma and by the TO-line of free excitons. The discrepancy between the theory and the experiment in the high energy tail results from absorption of the light. The spectral position of the low energy edge of the EH-line $h\nu_o$ gives a value of the silicon energy gap $E_g$, depending on the temperature. The temperature dependence of the silicon energy gap is represented by the formula [23,24]

$$E_g = E_g^o - \alpha_g T_g^2 (T_g + \beta_g)^{-1}, \qquad (2)$$

where $E_g^o = 1.17\,eV$ is the silicon energy gap at $T_g = 0$, $\beta_g = 636\,K$, $\alpha_g = 4.73 \times 10^{-4}\,eV\,K^{-1}$. Here we introduce the energy gap temperature $T_g$, representing the temperature of the lattice inside the observed electron-hole plasma. $T_g$ can be different from the lattice temperature $T$ outside the observed plasma. For the energy gap outside the electron-hole plasma the lattice temperature $T$ should be present instead of $T_g$ in the formula (2). An overheating of the diode by increasing diode current is accompanied a decrease of the semiconductor energy gap and a corresponding shift of the luminescence line to the low energy side of the spectrum (figure 3). This gives the dependence of the energy gap temperature $T_g$ on the diode current, that can be compared with the dependence of the spectral temperature $T'$ on the diode current (figure 2). The maximum value of the diode temperature in our experiments is $T \approx 350\,K$ for Si:P and Si:B diodes. The energy gap temperature, being quite different from the diode temperature (figure 2), indicates, that the temperature of the lattice inside the plasma is essentially higher than the temperature outside the plasma. This implies, that the energy gap inside the plasma is lower than the energy gap outside the plasma.

At the diode current $J° \geq 150\,mA$ the efficiency of the Si:P diodes is estimated as $\gamma\eta \approx (1 \div 3) \times 10^{-3}$. At the diode current $J° = 200\,mA$ the integral intensity of the EH-line is much higher than the integral intensity of electroluminescence of free excitons, observed in the same diodes at the temperature $T = 140\,K$ [14-16]. The energy gap temperature quite different from the diode temperature, the optical hysteresis in the intensity-current dependence and the very high quantum efficiency of the luminescence can be explained by condensation of injected electron-hole plasma in the Si:P diodes into a dense surface state. A strong reduction of the silicon energy gap due to generation of phonons by the plasma is a possible reason of the plasma condensation. The threshold optical hysteresis is a nucleation phenomenon characteristic of the threshold condensation of the electron-hole plasma. A similar threshold optical hysteresis was observed in the intensity of luminescence of electron-hole drops in thin samples of germanium under optical excitation [25-27]. At very low injection currents the diffusion-drift model seems to be valid. The existence of the optical hysteresis indicates a threshold origination of dense surface plasma drops or a drop with increasing diode current. It can be assumed, that at high diode currents the injected plasma forms a surface drop with a constant plasma density, a constant quantum efficiency and the radius increasing with increasing diode current. This could explain the long linear part in the intensity-current dependence [7]. In accordance with the formula (1) the density of the surface electron-hole plasma, corresponding to the observed quantum efficiency of the luminescence, is estimated as $n \approx (1 \div 3) \times 10^{18} cm^{-3}$. A detailed model of the plasma condensation is discussed in section "the model of the plasma condensation".

In the case of the Si:B diodes, two types of diodes are realized, type 1 (Si:B-1) and type 2 (Si:B-2) [7]. For the Si:B-2 diode the voltage-current dependence and the intensity-current dependence are bistable (figure 2). The voltage-current dependence consists of the upper current branch, observed at low diode currents, and the lower branch, observed at high diode currents. The intensity-current dependence consists of the intensity branch $I_+$, corresponding to the upper current branch, and the intensity branch $I_-$, corresponding to the lower current branch. Switching between current branches and simultaneous switching between intensity branches occur at the threshold currents $J_{t1}^o$ and $J_{t2}^o$ (figure 2). The luminescence intensity at the upper current branch is much lower than the luminescence intensity at the lower current branch. A well-defined hysteresis is observed in the current-voltage dependence of the Si:B-2 diode, and a corresponding optical hysteresis is observed in the intensity-current dependence (figure 2). At the lower current branch the EH-line is broad (figure 3), and the electronic temperature of the plasma is close to the energy gap temperature (figure 2). In addition to the EH-line a new $EH_s$-line arises in electroluminescence spectra with increasing diode current at the lower current branch (figure 3, curves $2_-$ and $2'_-$). The energy difference of the spectral positions of these lines is equal to the energy difference $\Delta = 44\ meV$ between the valence band $\Gamma_8$ and the valence band $\Gamma_7$, split off by the spin-orbit interaction. The $EH_s$-line is attributed to recombination radiation of free electrons from the indirect conduction band minima and free holes from the valence band $\Gamma_7$. At the upper current branch at low diode currents the EH-line is relatively narrow and corresponding electronic temperature of the plasma is low (figure 4, spectra $2_+$ and $2'_+$). In addition to the EH-line the $EH^+$-line is observed in the luminescence spectra. The energy difference of the spectral positions of these lines is equal to two energies of the TO-phonon $2\eta\Omega^{TO} = 116\ meV$. The $EH^+$-line is a replica of the EH-line, corresponding to absorption of the TO-phonon in the recombination process. The intensity ratio of these lines, increasing with increasing diode current, is equal to $N_q/(1+N_q)$, where $N_q$ the filling number of the TO-phonon. We have found, that the filling number of the TO-phonon is defined by the energy gap temperature and is equal to $N_q = [\exp(\eta\Omega^{TO}/kT_g)-1]^{-1}$. With increasing diode current the $EH_s$-line arises in the luminescence spectrum at the upper current branch due to occupation of the valence band $\Gamma_7$ by hot holes in a strong electrical field (figure 4, spectrum $2''_+$). A high mobility of holes in the valence band $\Gamma_7$, resulting from the low hole effective mass $m^s_{oh} = m^s_{dh} \approx 0,25\ m_o$ [22], could give an essential overheating of the electron-hole plasma and broadening of the luminescence line in a strong electrical field. The mobility of holes in the valence $\Gamma_8$ is low due to the high hole effective mass of density of states $m_{dh} = 0.58\, m_o$ [21-23].

For the Si:B-1 diode a negative differential resistance is observed in the voltage-current dependence (figure 5). The negative differential resistance arises at the diode currents higher than the threshold current, corresponding to the origination of the strong luminescence of injected plasma. The voltage-current dependence is reversible. At low diode currents the voltage-current dependence of the Si:B-1 diode reminds the upper current branch of the Si:B-2 diode. The intensity-current dependence reveals a well-defined threshold hysteresis. The luminescence intensity for increasing diode current (figure 5, curve $I_+$) is lower than the luminescence intensity for decreasing diode current (figure 5, curve $I_-$). At high diode currents the intensity-current dependence is practically linear. At the maximum diode current $J° = 200\, mA$ the efficiency of the Si:B-1 and the Si:B-2 diodes is $\eta \approx (1 \div 3) \times 10^{-3}$. This gives the plasma density in the diodes equal to $n \approx (1 \div 3) \times 10^{18}\, cm^{-3}$. The EH-line dominates in the luminescence spectrum of the Si:B-1 diode (figure 6), and the EH$^+$-line is present at the high energy side of the spectrum. The EH$_s$-line arises in the luminescence spectrum with increasing diode current, but its intensity is low in spite of the strong electrical field in the silicon substrate. The electronic temperature of injected plasma $T_e = \varsigma^{-1} T'$, obtained from the spectral position of the EH-line maximum $h\nu_m = h\nu_o + 2kT'$, is unusually low (figures 5 and 6), however, $T_e$ at the lower current for the Si:B-2 diode is close to the energy gap temperature. In any case the electronic temperature of the plasma should be higher than the temperature of the lattice due to an overheating of the plasma in a strong electrical field, and the low spectral temperature $T'$ now is unexplained.

The energy gap temperature quite different from the diode temperature, the negative differential resistance of the diodes, the optical hysteresis, the strong sensitivity of the $EH_s$-line intensity on the applied electrical field can be explained by condensation of injected electron-hole plasma into dense electron-hole flexes in the Si:B-1 diodes. A strong electrical field, producing a strong electrical current and an essential overheating of electrons and holes, can exist in dense electron-hole plasma only in the case of formation of dense electron-hole flexes or a flex with the length close to the substrate thickness. At strong recombination of electrons and holes the existence of dense plasma flexes is supplied by a current of electrons and holes from the substrate into the flexes. The recombination current inside a plasma flex should be equal to the current of electrons and holes into the flex. This condition gives the radius of the flex $R = 2v\tau_R(n/n_o)$. Here $v$ is the velocity of electron-hole pairs, defined by drift and diffusion of low-density plasma outside the flex, $\tau_R$ is the recombination time inside the flex, $n$ is the plasma density outside the flex, $n_o$ is the plasma density inside the flex. The optical hysteresis in the Si:B-1 diode is similar to the optical hysteresis in the Si:P diode and represents presumably a similar threshold nucleation phenomenon. At very low injection currents the diffusion-drift model seems to be valid in the diodes. The existence of the optical hysteresis indicates a threshold origination of dense plasma flexes or a flex with increasing diode current. It can be assumed, that at high diode currents in the region of the negative differential resistance of the Si:B-1 diode only a single plasma flex exists with almost constant plasma density and the radius increasing with increasing diode current. This could supply the long linear intensity-current dependence [7]. The decrease of the gate voltage with increasing diode current in the region of the negative differential resistance was tentatively explained by occupation of the valence band $\Gamma_7$ by hot high mobility holes in a strong electrical field [7]. A correct theory of this phenomenon is absent, and the role of holes from the valence $\Gamma_7$ should be verified. In any case hot dense plasma with a large fraction of holes from the valence band $\Gamma_7$ exists inside the plasma flex and low-density plasma in a drift regime is realized outside the flex. The plasma condensation is accompanied by a strong concentration of the drift and recombination currents inside the plasma flex.

The bistable voltage-current dependence and intensity-current dependence of the Si:B-2 diode with the luminescence spectra of the diode can be explained by the same way. At the upper current branch, unstable at high diode currents, origination of dense plasma flexes or a flex occurs with increasing diode current until the switching of the gate voltage into the lower current branch. A drift regime of low-density plasma with the recombination current, concentrated mostly at the $p^+$-contact, is realized at the upper current branch in a strong electrical field. At the lower current branch, unstable at low diode currents, presumably a single dense plasma flex with a large radius is formed. The radius of the flex decreases with decreasing diode current until the switching of the gate voltage into the upper current branch. A drift regime of the plasma outside the flex is also realized, but with the plasma density higher and the drift velocity lower than those at the upper current branch. The switching between these two branches with increasing and decreasing diode current, accompanied by the switching of the plasma distribution and the plasma luminescence, is a threshold phenomenon in the plasma condensation. Meantime, the electrical field at the lower current branch is low and we can not exclude completely, that a single plasma drop is formed presumably near the $p^+$-contact instead of a single plasma flex at the lower current branch. As it is shown below, the energy gap temperature inside such a drop should be maximum.

**The model of the plasma condensation**

A negative heat capacity of injected electron-hole plasma, concentration of the input diode power inside the plasma and weak diffusion of phonons at high temperatures represent the main physical reasons and conditions of the plasma condensation. Under these conditions generation of phonons by the plasma results in a local overheating of the lattice and a reduction of the semiconductor energy gap inside the plasma. In such a way the self-organized potential wells, attracting injected electrons and holes, is created (figures 1 and 7). The average energy of free electrons and holes in the electron-hole plasma in silicon under equilibrium between the plasma and the lattice is

$$E = E_g + 3kT_g = E_g^o - \alpha_g T_g^2 (T_g + \beta_g)^{-1} + 3kT_g. \qquad (3)$$

Here $E_g$ is the semiconductor energy gap, defined by the formula (2), $3kT_g$ is the average kinetic energy of electron-hole pairs at $T_e = T_g$. In a strong electrical field the kinetic energy is $3kT_e$. The heat capacity of the electron-hole plasma is $C_v^p = (3k + dE_g/dT_g)n$. In accordance with the formula (3) the heat capacity of electron-hole plasma in silicon is negative ($C_v^p < 0$) at $T_g > T_c$, where $T_c \approx 320\ K$ is the critical temperature. At temperatures higher than the critical temperature the average energy of the plasma decreases with increasing temperature and electron-hole pairs should be concentrated in a crystal region with the highest temperature. At a low injection potential $U'$ (figure 1) and the plasma density higher than the impurity concentration the main part of the diode input power is transferred to the electron-hole plasma and from the plasma is transferred to the lattice. Hence, the injected plasma is the main source of the lattice overheating in the diodes. The density of the phonon energy is generated by the plasma with the generation rate

$$G = [(E_g + 3kT_g)\tau_R^{-1} + e(\mu_e + \mu_h)\mathcal{E}^2]\, n. \qquad (4)$$

Here the first term results from the recombination of electrons and holes and the second term represents the Joule heat. The distribution of the lattice temperature in the sample is described by the heat conductivity equation

$$div(\kappa\, gradT_g) + G = 0, \qquad (5)$$

where $\kappa$ is the coefficient of heat conductivity. At room temperature this coefficient is equal to $\kappa = 1.5\, W\, cm^{-1} K^{-1}$, and the diffusion coefficient of phonons $D_a = 0.9\, cm^2 s^{-1}$ is lower than the diffusion coefficient of low-density plasma $D \approx 10\, cm^2 s^{-1}$ [23]. The coefficient of heat conductivity decreases with increasing temperature and increases with increasing plasma density [21,23]. If we assume, that the electron-hole plasma is concentrated in a flex with the radius $R$, as it is shown in the figure 7, the energy gap temperature outside the flex decreases with increasing distance $r$, and the extra temperature inside the flex, derived from the equation (5), is

$$T_g - T \approx (G/\kappa)\, R^2, \qquad (6)$$

where $T$ is the temperature at the distance $r = 2R$ from the boundary of the flex, $T_g$ is the temperature inside the flex. Boundary conditions for the heat flow depend on a cooling regime of the diode and we assume for the simplicity, that the temperature $T$ in the formula (6) is close to the diode temperature. The same solution for the extra temperature is obtained for a spherical plasma drop and for a plasma layer with $R$ equal to the drop radius or the layer thickness. The extra temperature depends on the plasma density and on the total number of carriers in the plasma. The formula (3) with the formulae (4) and (6) show, that at $T > T_c$ the average energy of carriers decreases with increasing plasma density. In the case of a spherical plasma drop this indicates an attraction between carriers changed by the Fermi repulsion at very high plasma densities. Simple calculations by use of these formulae result in the surprising conclusion: the energy of a spherical plasma drop with a high fixed total number of carriers should reach a minimum at the plasma density $n_o > 10^{18} cm^{-3}$. This conclusion is valid for a plasma flex and a plasma layer, if Auger recombination is sufficiently strong. Therefore, at $T > T_c$ the electron-hole plasma in silicon creates an attractive field producing the self-compression of the plasma. Estimates of the energy gap temperature for a plasma flex in the Si:B-1 diode at the maximum diode current agree with the experiment at $n_o \approx 3 \times 10^{18} cm^{-3}$ and $R \approx 10^{-3} cm$. The strong overheating of the plasma is achieved due to weak diffusion of phonons at high temperatures and a long lifetime of the extra phonons $\tau_a \approx R^2 / D_a$. In a surface plasma layer the extra temperature should be low due to low Joule heat. The extra temperature reaches a maximum value, if a large single drop is formed in the substrate. Condensation of injected plasma into such a drop at the silicon surface in the Si:P diode should supply the observed overheating of the plasma.

So, our simple model gives the satisfactory description of the experiments. A rigorous solution of the problem can be obtained from a system of equations including the continuity equations, the Poisson equation and the heat conductivity equation. This solution should describe behavior of a coupled plasma-phonon system in an electrical field at high temperatures. The plasma condensation seems to be a universal phenomenon in semiconductors with a negative heat capacity of electron-hole plasma.

**Conclusion**

We have observed the unusual phenomenon – high temperature condensation of injected electron-hole plasma in silicon. Origination of this phenomenon is explained by the plasma-phonon coupling, resulting from the strong modulation of the semiconductor energy gap under generation of phonons by the plasma at high temperatures. Our experiments give not sufficient arguments to identify the plasma condensation as a phase transition. The plasma condensation in silicon is similar to the compression of star matter under gravitation and can be identified as the self-compression of injected electron-hole plasma. The high quantum efficiency of the plasma luminescence in the diodes gives an opportunity for realization of silicon optoelectronics.

**Figure captions**

**Figure 1.** Band diagram of the Si:P diode in the model of the plasma condensation.

**Figure 2.** Dependence of the gate voltage $V_g$ ($V_g, V_g^+, V_g^-$), the electroluminescence intensity $I$ ($I, I_+, I_-$), the energy gap temperature $T_g$ ($T_g, T_g^-$) and the spectral temperature $T'$ ($T', T'^-$) on the diode current $J°$ for the Si:P diode and for the Si:B-2 diode; $I_-, T_g^-, T'^-$ correspond to the lower current branch with the voltage $V_g^-$; $I_+$ corresponds to the upper current branch with the voltage $V_g^+$; $I$ is the intensity of the luminescence line maximum.

$T \approx 293\ K$ at $J° = 0$ and $T \approx 350\ K$ at $J° \approx 200\ mA$.

**Figure 3.** Electroluminescence spectra of the Si:P diode $(1, 1')$ and the Si:B-2 diode at the lower current branch $(2_-, 2'_-)$. The dashed lines represent the theory.

$1: J° = 30.2\ mA, V_g = 4.2\ V, T_g = 410\ K, T' = 350\ K$.

$1': J° = 200\ mA, V_g = 13\ V, T_g = 500\ K, T' = 410\ K$.

$2_-: J° = 110\ mA, V_g = -10.7\ V, T_g = 460\ K, T' = 460\ K$.

$2'_-: J° = 200\ mA, V_g = -13.4\ V, T_g = 500\ K, T' = 500\ K$.

**Figure 4.** Electroluminescence spectra of the Si:B-2 diode at the upper current branch. The dashed lines represent the theory.

$2_+: J° = 100\ mA, V_g = -12.2\ V, T_g = 460\ K, T' = 300\ K$.

$2'_+: J° = 120\ mA, V_g = -16\ V, T_g = 480\ K, T' = 310\ K$.

$2''_+: J° = 148\ mA, V_g = -26\ V, T_g = 520\ K, T' = 580\ K$.

**Figure 5.** Dependence of the gate voltage $V_g$, the electroluminescence intensity $I$ $(I_+, I_-)$, the energy gap temperature $T_g$, the spectral temperature $T'$ on the diode current $J°$ for the Si:B-1 diode. $I$ is the intensity of the luminescence line maximum. $T \approx 293\ K$ at $J° = 0$ and $T \approx 350\ K$ at $J° \approx 200\ mA$.

**Figure 6.** Electroluminescence spectra of the Si:B-1 diode. The dashed lines represent the theory.

$3: J° = 100\ mA, V_g = -25.2\ V, T_g = 520\ K, T' = 330\ K$.

$3': J° = 140\ mA, V_g = -35.4\ V, T_g = 600\ K, T' = 350\ K$.

$3'': J° = 200\ mA, V_g = -29.4\ V, T_g = 610\ K, T' = 330\ K$.

**Figure 7.** The distribution of the plasma density $n$, the extra temperature $T_g - T$ and the semiconductor energy gap $E_g$ in the electron-hole plasma flex or in the electron-hole plasma drop in silicon (qualitative behavior).

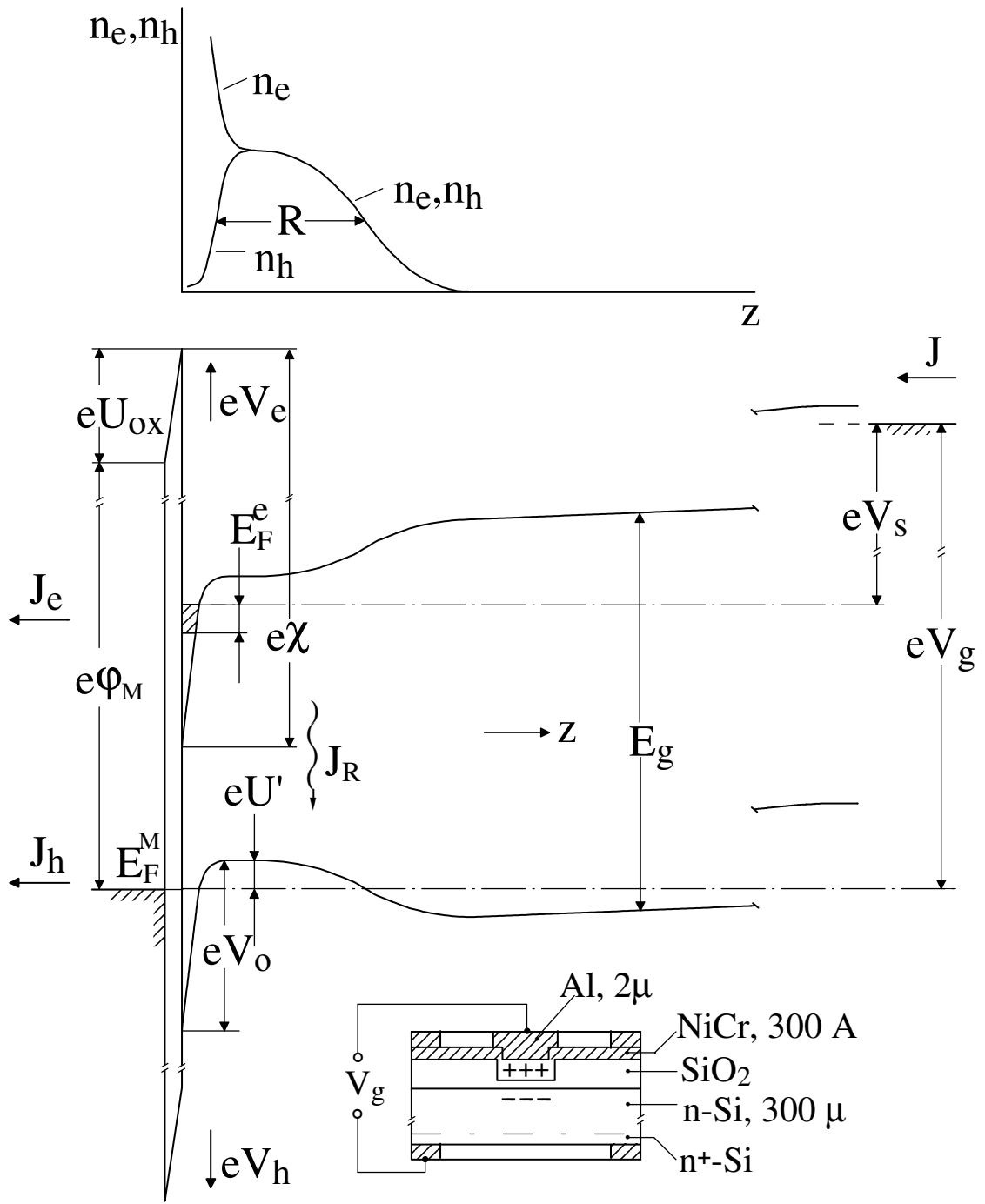

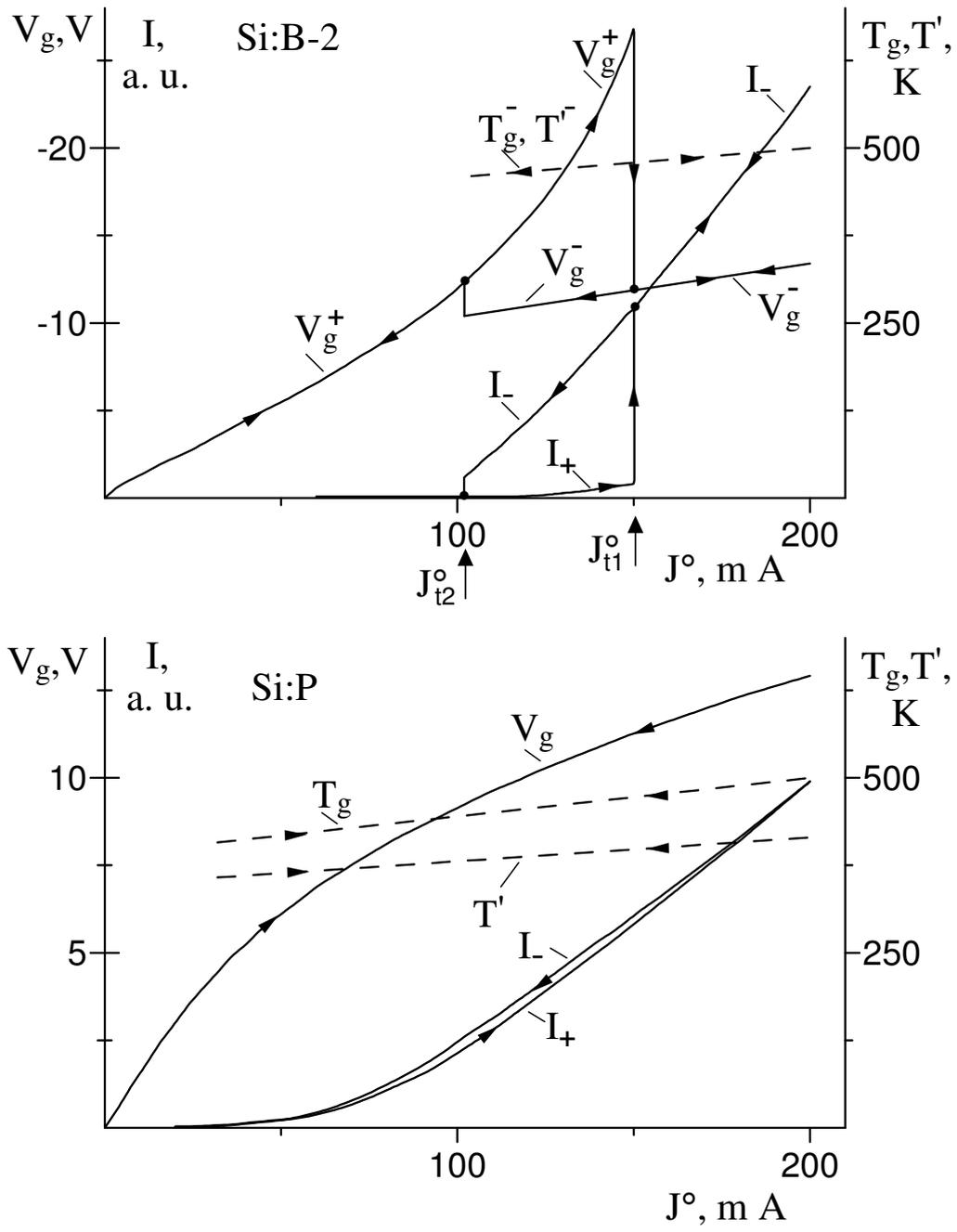

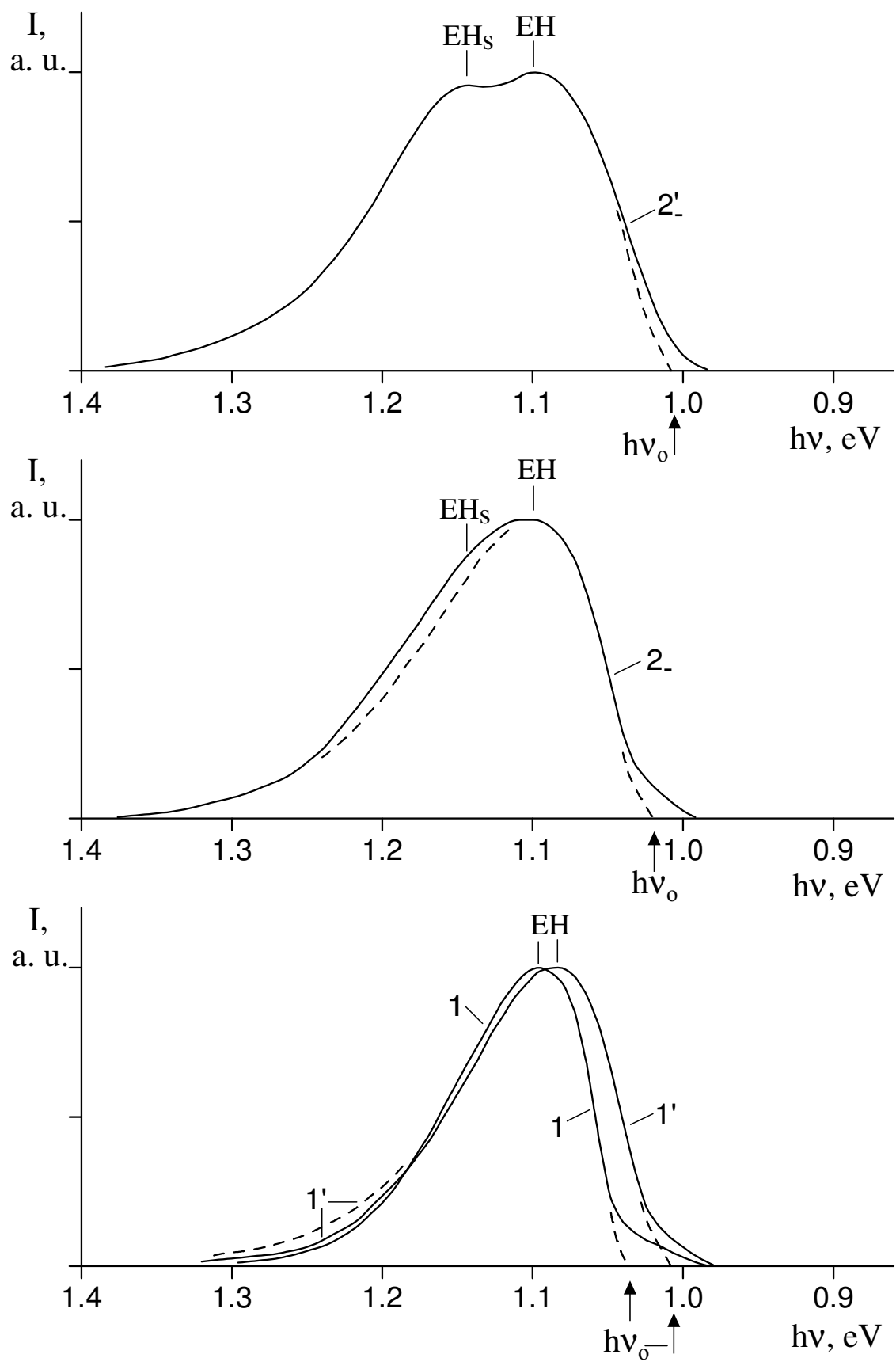

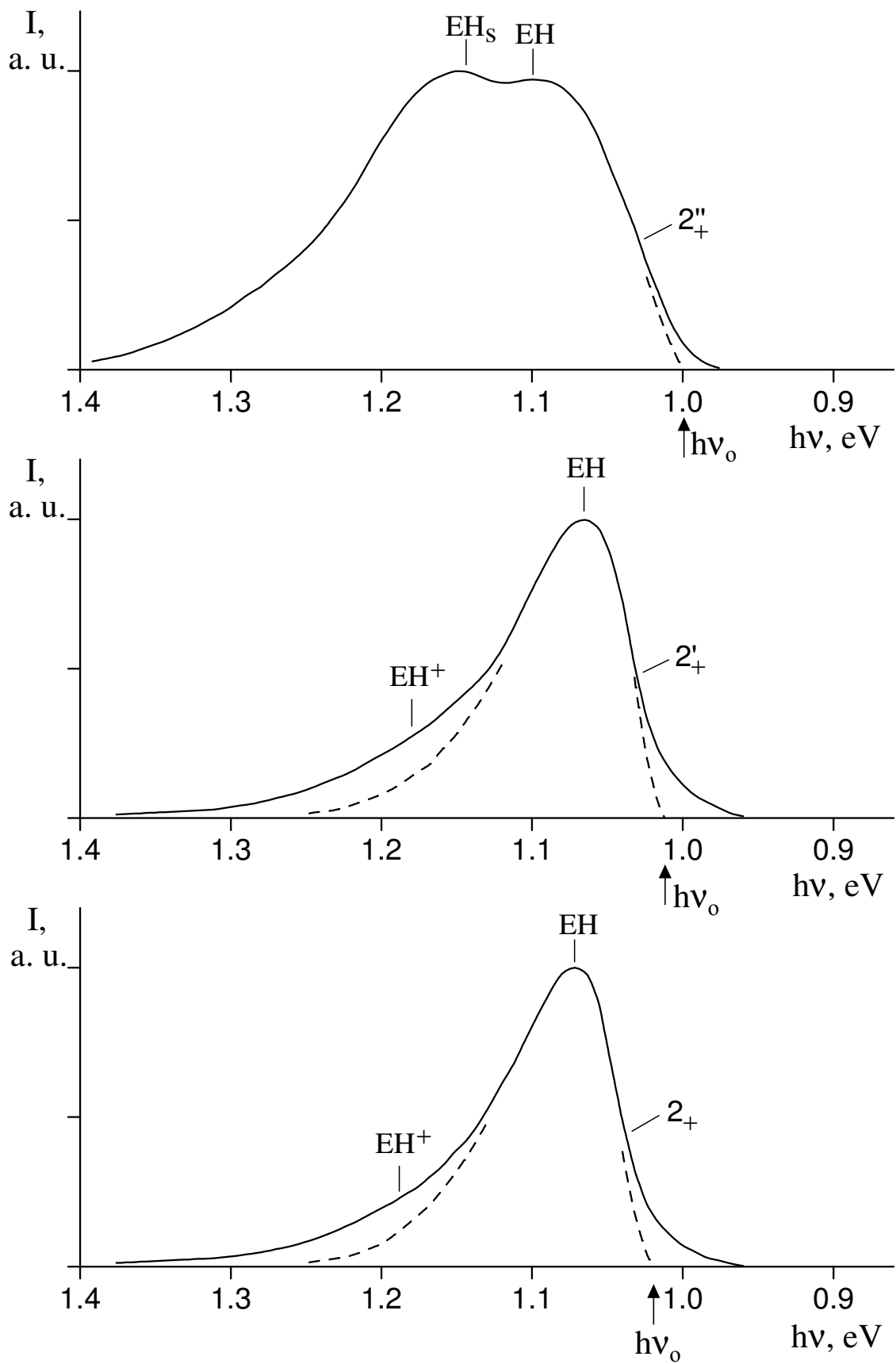

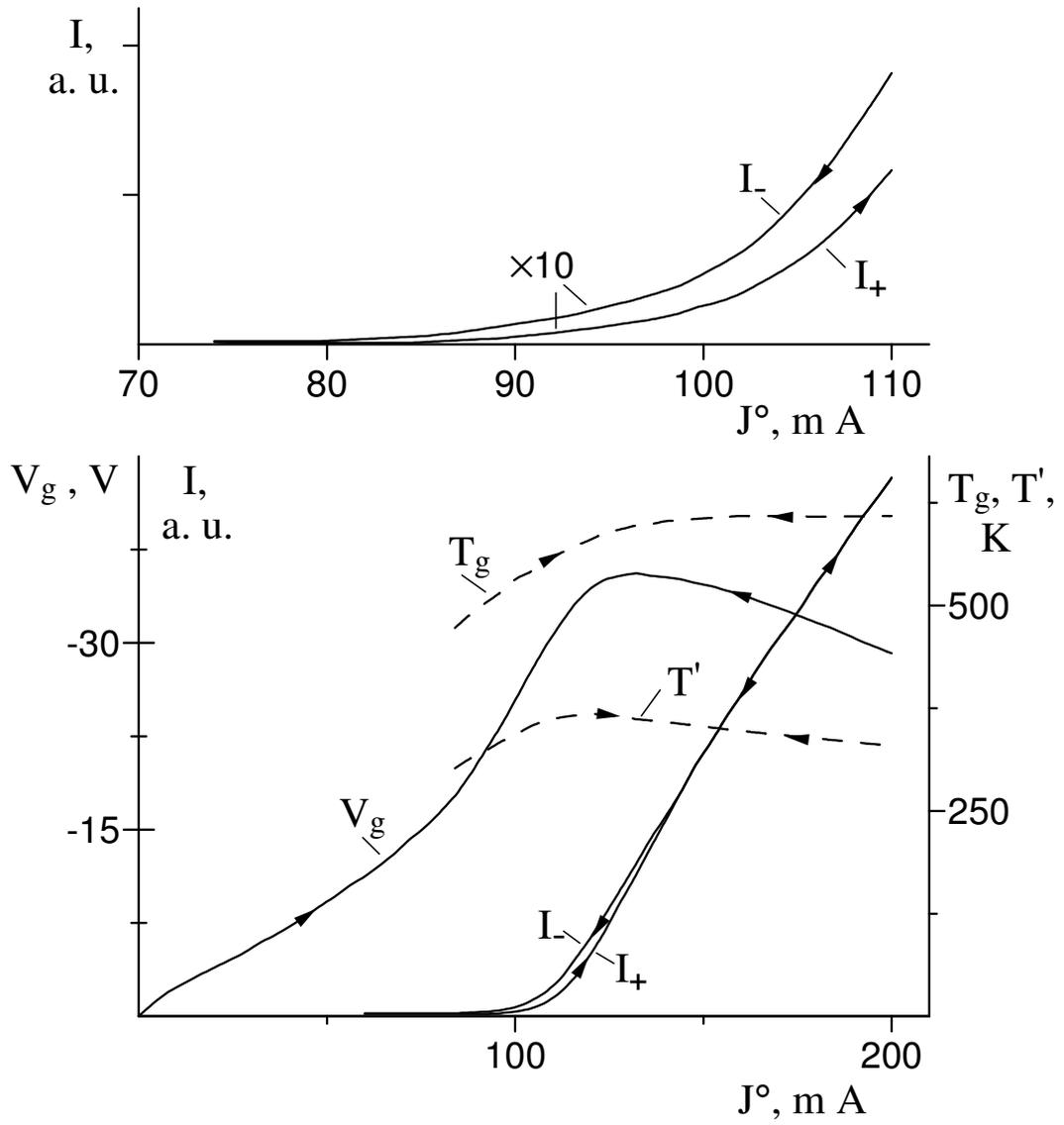

Figure 6

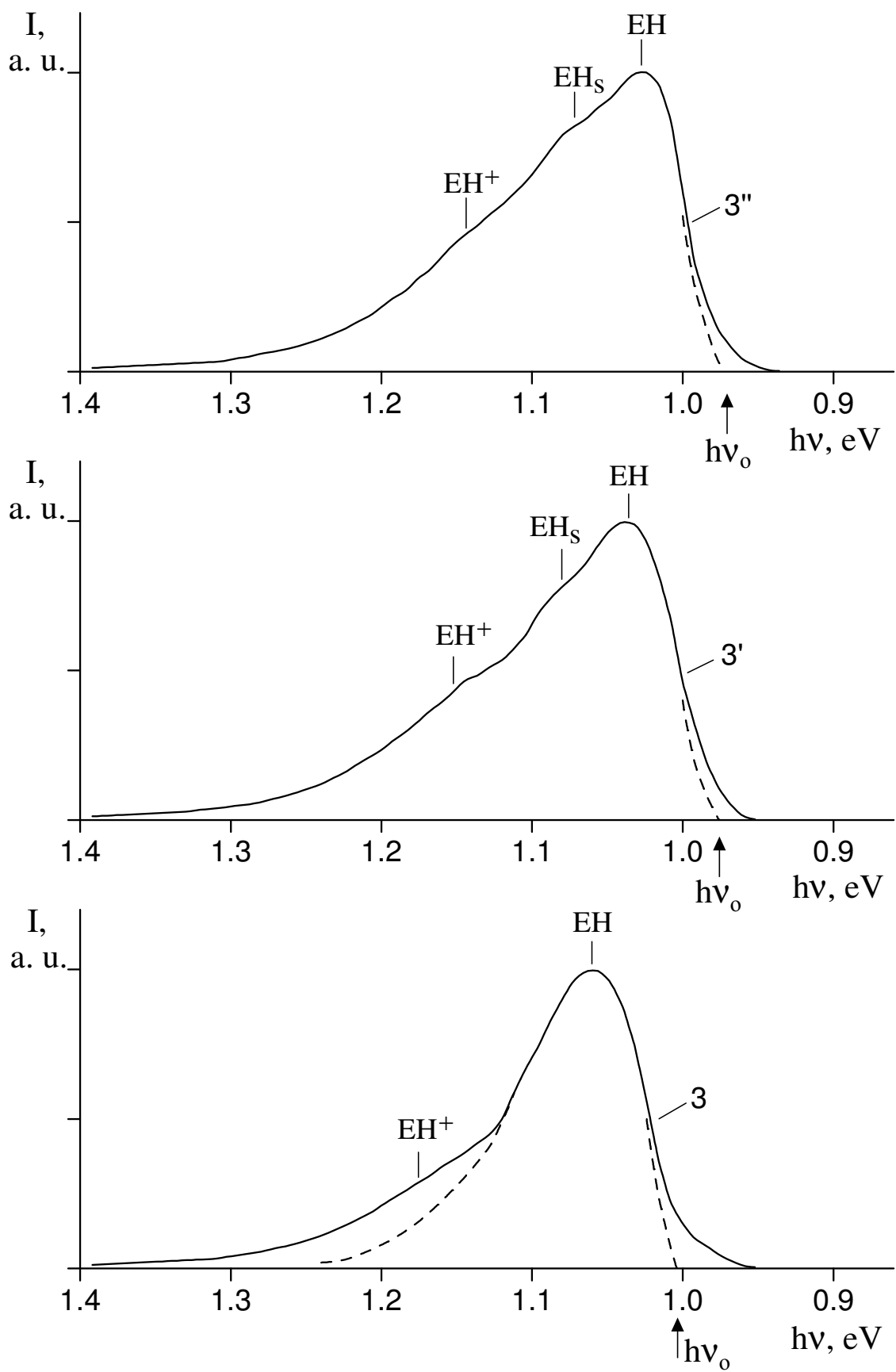

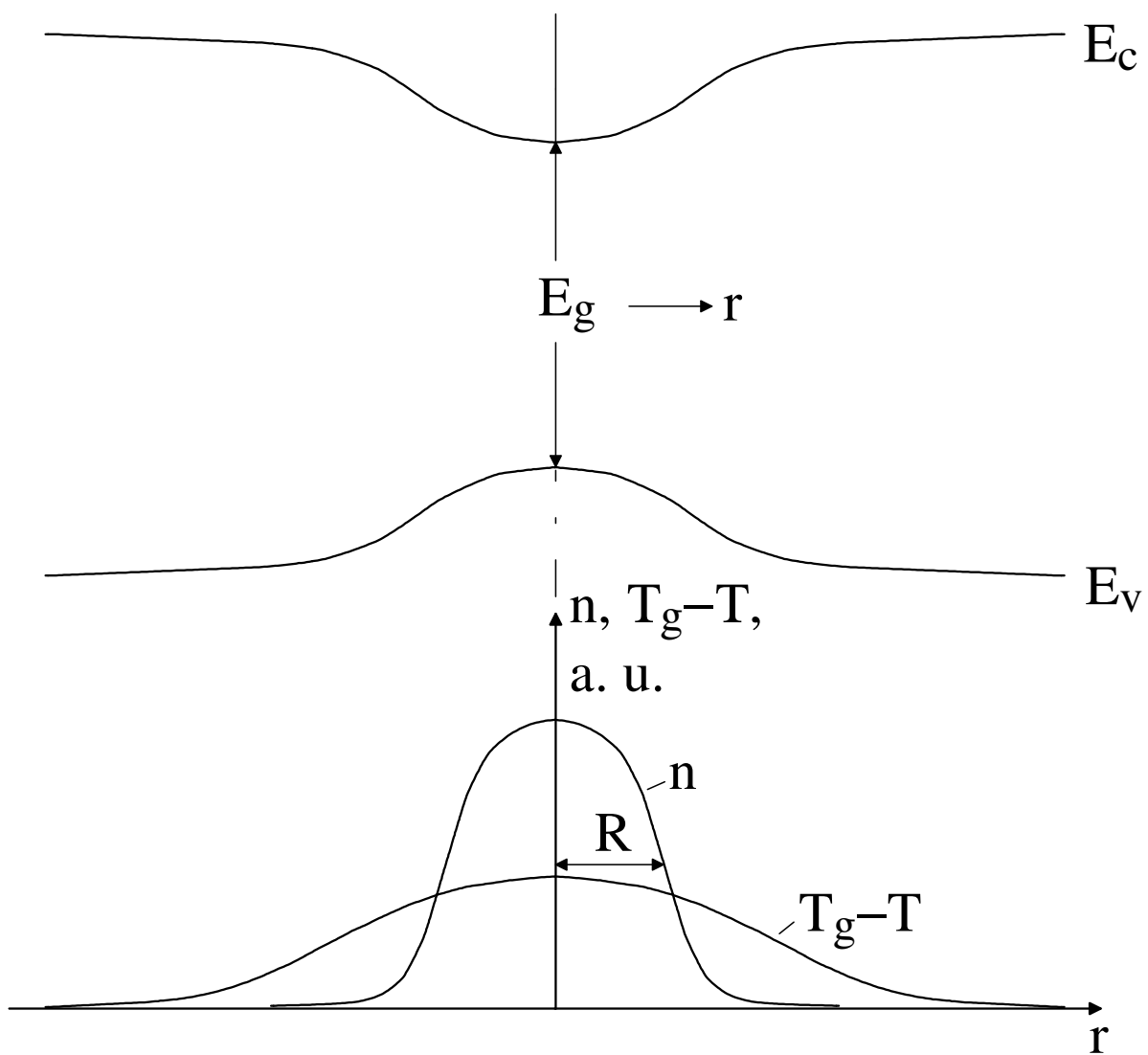